\documentstyle[preprint,aps,prl,epsf]{revtex}
\begin{document}
\title{Decoherence at absolute zero}
\author{Supurna Sinha\footnote{E-mail: supurna@rri.res.in}}
\address{Raman Research Institute. Bangalore 560 080. India}
\maketitle
\widetext

\begin{abstract}
We present an analytical study of the loss of quantum coherence at absolute 
zero. Our model consists of a harmonic oscillator coupled to an 
environment 
of harmonic oscillators at absolute zero. We find that for an Ohmic bath, 
the offdiagonal elements of the density matrix in the position 
representation decay as a power law in time at late times. This slow loss of 
coherence in the quantum domain is {\it qualitatively} different from the 
exponential decay observed in studies of high temperature environments.\\~\\
\end{abstract}

\section{Introduction}

\noindent
Quantum mechanics gives an excellent description of the microscopic world. 
In all experimental situations studied so far, the predictions of quantum 
theory have been verified with phenomenal accuracy. However, ever since the 
theory came into being it has caused a deep unease among physicists. This 
dissatisfaction originates from a belief that the foundations of the theory 
are infirm, or at best poorly understood. In order to verify the predictions 
of quantum mechanics, one has to perform {\it measurements}. However, 
quantum mechanics does {\it not} provide a description of the measuring 
process itself. Apart from a dynamical description of the quantum system 
via Schr\"{o}dinger's equation, one needs to postulate [1] {\it ad hoc}
that the state of the system ``collapses" into one of the possible 
classical outcomes selected by a measuring apparatus. This point of view, 
which is popularly known as the Copenhagen interpretation, 
explicitly requires the existence of a classical measuring apparatus. It is 
clearly unsatisfactory that for its interpretation quantum mechanics needs 
to rest upon classical mechanics, which it is supposed to supersede. A more 
general theory usually subsumes a less general one as for instance, the 
special theory of relativity contains Newtonian physics as a special case. 
One can formulate the special theory of relativity independent of Newtonian 
physics. The situation is much less satisfactory for quantum mechanics. The 
basic problem stems from the superposition principle, which is one of the 
corner stones of quantum mechanics. This principle states that if two states 
of a system are allowed, an arbitrary linear combination of them also is. As 
a consequence, these two allowed states can occur with a definite phase 
relation. It is this ``coherent superposition" of alternatives which 
is in apparent conflict with everyday experience. Coherent superposition 
gives rise to the possibility of interference between alternatives, in some 
sense that both alternatives simultaneously occur. This is best illustrated 
by Schr\"{o}dinger's cat paradox. One can devise a situation in which the life 
of a cat depends on whether a radioactive nucleus has decayed. This makes 
the state of a macroscopic object (the cat) depend on the state of a 
microscopic one (the nucleus). If the state of the nucleus is in a 
superposition of two states (decayed and not decayed), the cat too will have 
to be in a superposition of two states - alive and dead. This thought 
experiment brings out the absurd nature of quantum mechanics, when it is 
applied to the macroscopic world. This is sometimes referred to in the 
physics community as ``the measurement problem" [2]. There have been many 
attempts to address this problem, but unfortunately, with very little 
progress. \\

\noindent
In the last few decades, a fresh perspective has emerged on this issue [3]. 
This new point of view uses quantum Brownian motion as a model for 
understanding the loss of quantum coherence due to a system's contact with 
its environment. This interest has been triggered by the possibility of 
probing the boundary between the microscopic and the macroscopic worlds via 
increasingly sophisticated experiments [4]. In this open-system approach to 
quantum measurement, one models the quantum system as a free particle in a 
potential $V(x)$ and its environment as a collection of harmonic oscillators 
[5,6]. The effective dynamics of the quantum system is then studied by 
``ignoring" the environmental degrees of freedom. More precisely, the 
dynamics of the reduced density matrix (which is obtained from the full 
density matrix of the system plus the environment by tracing out the 
environmental degrees of freedom) is analyzed. This approach leads to the 
conclusion that quantum coherence between two Gaussian wave-packets 
separated by a distance $\Delta x,$ decays exponentially over a time-scale 
known as the decoherence time 
$\tau_D = \gamma^{-1} (\lambda_{\rm Th}/\Delta x)^{2}$, 
where $\lambda_{\rm Th}$ is the thermal de Broglie 
wavelength and $\gamma$ the relaxation rate of the system. In other words, 
a superposition of states cannot be observed if the system is probed at time 
scales greater than $\tau_D$. Thus one seems to get an explanation for the 
loss of quantum coherence {\it within} the framework of quantum mechanics. 
However, most of these studies take the high temperature 
$(T \rightarrow \infty)$ limit for analytic convenience. 
While such an assumption leads to some technical simplifications it is 
inadequate from a conceptual point of view. Since the high temperature limit 
($T \rightarrow \infty$) is equivalent to the classical limit 
($\hbar \rightarrow 0$) such an analysis seems to require classical mechanics 
for understanding the decoherence aspect of 
quantum mechanics just as in the Copenhagen interpretation.\\

\noindent
In this paper we address and partially answer the following questions: Is 
there a loss of quantum coherence even at absolute zero? If so, what is 
the law of decoherence in the absence of thermal fluctuations? The main 
motivation for this study is the following. Since an environment at absolute 
zero is at the extreme quantum limit (where the analysis simplifies), the 
present study enables us to understand the process of loss of quantum 
coherence entirely {\it within} the framework of quantum mechanics. 
This is a conceptually important point which sets the present work apart 
from most earlier attempts towards understanding the measurement aspect 
of quantum mechanics via environment induced decoherence. Unruh and 
Zurek [7], Paz, Habib and Zurek [8] and Caldeira and Leggett [9] 
have addressed similar questions in the past. However, the work of 
Unruh and Zurek [7] is based on an incorrect regularization procedure which 
is in conflict with the fluctuation-dissipation theorem. At the end of this 
paper we discuss how the work discussed here relates to Refs. [8,9].\\

\noindent
We consider a harmonic oscillator coupled to an environment characterized 
by an Ohmic spectral density 
(i.e ($I(\omega) = (2/\pi) \Gamma\omega\theta (\Lambda - \omega $), 
with $\Gamma$ the friction coefficient and $\Lambda$ the 
upper cut-off on the frequency spectrum of the environmental oscillators). 
The environment is coupled to the system via a coordinate- coordinate 
coupling. In the limit of weak coupling between the system and the 
environment, we find that the quantum system exhibits a new behavior for the 
loss of quantum coherence at absolute zero.\\

\section{Evolution of the reduced density matrix}

\noindent
There have been several attempts at deriving a general master equation for a 
quantum system coupled to a harmonic oscillator environment [5-7,10]. One of 
the first attempts in this direction was due to Caldeira and Leggett [6] 
who gave an explicit derivation of the evolution of the reduced density 
matrix valid for a high-temperature environment. Since their work, there 
have been many derivations under more and more general conditions [7,10]. 
Unfortunately, many of these derivations are internally inconsistent. As 
mentioned earlier, the derivation of Unruh and Zurek [7], which is supposed 
to be valid for an Ohmic environment at all temperatures makes use of an 
incorrect regularization procedure which violates the 
fluctuation-dissipation theorem. The most general and correct derivation so 
far has been due to Hu, Paz and Zhang (HPZ) [10]. HPZ have incorporated all 
the subtleties which come into play at low temperatures. These subtleties 
stem from the {\it nonlocality} of the noise kernel at low temperatures.\\

\noindent 
In this paper we base our analysis on the master equation derived by HPZ. 
The Hamiltonian for the composite system described by HPZ is given by 

\begin{equation}
H = H_{S} + H_{I} + H_{E},
\end{equation}

with $H_{S}$ and $H_{E}$ representing the Hamiltonian of the system and the 
environment respectively. $H_{I}$ is the Hamiltonian which pertains to 
interaction between the system and the environment. The explicit 
expressions for $H_{S}, H_{I}$ and $H_{E}$ are 

\begin{eqnarray*}
H_{S} &=& \frac{P^{2}}{2M} + \frac{M\Omega^{2} x^{2}}{2},\;\;\;\;H_{I} = \sum_{n} C_{n}xq_{n}, \\
H_{E} &=& \sum_{n} \left( \frac{p^{2}_{n}}{2m_{n}} + \frac{m_{n}\omega^{2}_{n} q^{2}_{n}}{2}\right ).
\end{eqnarray*}

\noindent
Here $M, x$ and $P$ are respectively the mass, position and the momentum of 
the system while $m_{n}, q_{n}$ and $P_{n}$ are respectively the mass, 
position and momentum of the {\it n}th environmental oscillator. 
$\Omega$ and $\omega_{n}$ are the frequency of the system oscillator and 
the {\it n}th environmental oscillator respectively. $C_{n}$ is the coupling 
constant between the system and the {\it n}th environmental oscillator. 
The system and the environment are assumed to be decoupled initially. 
At $t = 0$ interaction is turned on. The corresponding evolution equation 
for the reduced density matrix for the quantum system in the coordinate 
representation is 

\begin{eqnarray}
i\hbar \frac{\partial\tilde{\rho} (x, x^{\prime}, t)}{\partial t} &=& 
\left[
- \frac{\hbar^{2}}{2M} \left( \frac{\partial^{2}}{\partial x^{2}} - 
\frac{\partial^{2}}{\partial x^{\prime 2}}\right)\right. \nonumber\\*[0.4cm]
&+& \left. \frac{M}{2} \Omega^{2} (x^{2} - x^{\prime 2})\right] 
\tilde{\rho} (x, x^{\prime}, t)\nonumber\\*[0.4cm]
&+& \frac{M}{2} \delta\Omega (t)^{2} (x^{2} - x^{\prime 2}) \tilde{\rho} 
(x, x^{\prime}, t) \nonumber\\*[0.4cm]
&-& i\hbar\Gamma (t)(x - x^{\prime}) \left(\frac{\partial}{\partial x} - 
\frac{\partial}{\partial x^{\prime}}\right) \tilde{rho}(x, x^{\prime}, 
t)\nonumber\\*[0.4cm]
&-& iD(t) (x - x^{\prime})^{2} \tilde{\rho} (x, x^{\prime}, 
t)\nonumber\\*[0.4cm]
&+& \hbar A(t) (x - x^{\prime}) \left( \frac{\partial}{\partial x} + 
\frac{\partial}{\partial x^{\prime}}\right) \tilde{\rho}(x, x^{\prime}, 
t).
\end{eqnarray}

\noindent
The terms in the square brackets make up the Liouvillian. The second term 
is a time dependent frequency shift of the system oscillator as a result of 
its coupling with the environment. The third term represents dissipation 
with a time dependent dissipative coefficient. The fourth term describes 
diffusion. The last term in the equation is named ``anomalous diffusion" in 
the literature since it generates a second derivative term in the phase 
space representation of the evolution equation just like the ordinary 
diffusion term. The time dependence of the coefficients entering this 
equation is in general rather complicated [10]. However, one can get simpler 
functional forms for these coefficients in some limiting cases. We will 
write down the explicit forms of the coefficients relevant to the present 
problem later in this paper. \\

\noindent
The evolution of the system involves the following time scales: the 
scale associated with the natural frequency $\Omega$ of the system 
oscillator, the relaxation time scale $\gamma^{-1}$ which is controlled 
by the coupling strength between the system and the environment, a ``memory 
time", which is the inverse of the highest frequency $\Lambda$ of the 
reservoir and the thermal time scale $\tau_{\beta} = \hbar\beta$ (with
$\beta = 1/k_{\rm B}T$ where $k_{\rm B}$ is the Boltzmann's constant 
and $T$ the temperature of the bath). In this paper we focus on the 
time regime $t \gg \tau$ where $\tau = \gamma^{-1}$ is the typical 
response time of the system. In other words, we are interested in the 
dynamics of loss of coherence at long times after some initial transients  
have died out.\\

\noindent
We now study how an initial coherent superposition of two wave-packets 
evolves into an incoherent mixture at an arbitrary temperature. In order to 
make analytical progress we need to make some reasonable approximations. For 
large enough separation (more precisely for 
$\vert(x - x^{\prime})\vert \gg \lambda_{q}$ where 
$\lambda_{q} = \sqrt{\hbar / M\gamma)}$\footnote{In the high temperature limit, 
one finds that coherence between two wavepackets is lost if they are 
separated by
a distance greater than the thermal de Broglie wavelength $(h/\sqrt{2Mk_{B}T)}.$
A naive extension of this formula to the low temperature regime would imply
that the coherence length goes to infinity, i.e. there is no destruction of 
coherence at {\it any} separation. Our analysis of the extreme quantum domain
shows that at zero temperature the coherence length is {\it finite} and given 
by $\sqrt{\hbar /M\gamma}$.} between the constituent wavepackets one can approximate 
Eq. (2) as follows, 

\begin{equation}
i\hbar \frac{\partial\tilde{\rho} (x, x^{\prime}, t)}{\partial t} \simeq 
-D(t)(x - x^{\prime})^{2} \tilde{\rho} (x, x^{\prime}, t).
\end{equation}

\noindent
This leads to the conclusion that the off-diagonal elements of the density 
matrix in the position representation get suppressed according to 

\begin{eqnarray}
\tilde{\rho} (x, x^{\prime}, t) &=& \tilde{\rho} (x, x^{\prime}, 
0)\nonumber\\
&& \times \exp \left( - \frac{(x - x^{\prime})^{2}}{\hbar}
\int^{t}_{0} D(t^{\prime} {\rm d}t^{\prime}\right).
\end{eqnarray}

\noindent
In the high temperature regime, Eq. (4) leads to the usual exponential 
damping of the off-diagonal elements of the density matrix [3]. In the 
limit of weak coupling between the system and the environment (more 
precisely, if one works up to the second order in the coupling constant 
between the system and the environment) one gets [10] 

\begin{equation}
D(t) = \int^{t}_{0} \nu (s)\cos \Omega s{\rm d}s,
\end{equation}
where $\nu(s)$ is the noise kernel given by 
\begin{equation}
\nu (s) = \int^{\infty}_{0} I (\omega) \coth \frac{\beta\hbar\omega}{2}
\cos \omega s {\rm d}\omega.
\end{equation}
Furthermore we restrict ourselves to the case of an Ohmic environment 
characterized by a spectral density $I(\omega) = (2/\pi)\Gamma\omega\theta
(\Lambda - \omega)$, with $\Gamma$ the friction coefficient 
and $\Lambda$ the upper cut-off on the frequency spectrum of the 
environmental 
oscillators. Such an environment leads to the classical law of friction in 
the limit of high temperatures. \\

\noindent
Thus for an Ohmic environment Eq. (6) goes over to 
\begin{equation}
\nu(s) = (2/\pi) M\gamma \int^{\Lambda}_{0} \omega \coth \frac{\beta\hbar\omega}{2} \cos \omega s{\rm d}\omega,
\end{equation}
where the dissipative coefficient $\gamma = \Gamma/M$. \\

\noindent
The explicit form of the integral appearing in the argument of the 
exponential in Eq. (4) at absolute zero is 
\begin{equation}
(2/\pi) M\gamma \int^{t}_{0} \int^{t^{\prime}}_{0} \int^{\Lambda}_{0}
\omega\cos \omega s \cos \Omega s{\rm d}s {\rm d}\omega.
\end{equation}
If we confine ourselves to the regime $t$ large compared to the 
relaxation time $\tau$ and the memory time $\Lambda^{-1}$ but 
still small compared to $\Omega^{-1}$, where $\Omega$ is the 
infrared cutoff\footnote{In this time regime the system essentially 
behaves like a free particle as it does not ``see'' the confining 
potential.}, Eq. (8) reduces to 
\begin{equation}
(2/\pi) M\gamma \int^{t}_{0} \frac{1 - \cos\Lambda t^{\prime}}{t^{\prime}} 
{\rm d}t^{\prime}.
\end{equation}
We find that the contribution from the rapidly oscillating cosine term is 
negligible compared to that from the first term in (9) as 
$\Lambda \rightarrow \infty$. We thus arrive at the following time
dependence for the off-diagonal elements of the reduced density 
matrix at late times (i.e. $\Omega^{-1} > t \gg \tau > \Lambda^{-1})$,
\begin{equation}
\tilde{\rho} (x, x^{\prime}, t) \sim t^{-\alpha},
\end{equation}
where $\alpha = (2/\pi\hbar) M\gamma (x - x^{\prime})^{2}$.

\noindent
This is the central result of the paper. We notice that, in contrast 
to the high temperature case, there is {\it no single characteristic 
decoherence time scale at absolute zero.} The slow power-law loss 
of quantum coherence at absolute zero is 
{\it qualitatively} different from the rapid suppression of 
coherence that one sees at high temperatures. This extremely slow 
decoherence stems from the {\it finite memory} in the system 
at absolute zero. More precisely, it can be traced to the temporal 
nonlocality of the noise kernel. The exponent $\alpha$ appearing in the 
decay of the off-diagonal elements of the density matrix is independent 
of the temperature reflecting the essential {\it quantum} nature of 
the dynamics.\\

\noindent
Let us discuss some of the consequences of this result. First of all it is 
intriguing that even at absolute zero a coherent superposition of states 
can get destroyed by zero point fluctuations. In other words, one is led to 
an understanding of the transition of an initial pure state into an 
incoherent mixture of states {\it within} the realm of quantum theory. 
Secondly, since this loss of coherence is very slow, we find that the 
effect of quantum zero point fluctuations on decoherence is considerably 
weaker than that of thermal fluctuations. There is one important 
difference between decoherence induced by a high temperature 
environment and that by a low temperature one. At absolute zero the 
quantum system can only lose energy to the cold environment. 
In contrast at higher temperatures the environment 
can also induce transitions between energy levels available to the quantum 
system. We would like to emphasize that a zero temperature environment is an 
extreme quantum system. Such an environment is ideal for an entirely quantum 
mechanical analytical approach towards the decoherence process. This fact 
has not been adequately appreciated in those earlier approaches towards the 
decoherence problem which are based on a high temperature (i.e. classical) 
limit for the environment. In the short time regime $(\gamma t \ll 1)$ 
studied by Caldeira and Leggett [9] one finds that dephasing occurs over 
a time scale of the order of the time for spontaneous emission of one 
quantum of energy 
from the system to the environment at absolute zero. On the other hand, at 
long times one can have spontaneous emission from a range of excited states 
of the system and consequently the decoherence process will be controlled by 
a wide range of time scales instead of a single time scale. This is 
perhaps the origin of the power law behavior of the decoherence process at 
large time scales at absolute zero. It would be interesting to find out if 
this slow loss of coherence is observed in experiments performed at 
cryogenic temperatures [4].\\

\noindent
There are several open questions which are of interest. An analytical 
study of the intermediate temperature regime would enable us to probe the 
competition between quantum and thermal fluctuations. In the present paper 
we have restricted our attention only to the case of an Ohmic environment. A 
similar derivation with a super-Ohmic or a sub-Ohmic environment may give 
us different dynamics for the suppression of the off-diagonal elements of 
the density matrix. \\

\noindent
In order to keep our analysis as simple as possible I we have confined 
ourselves to an approximate master 
equation (Eq. (3) ). In particular, here we have worked in a domain where 
the dynamics of the density matrix is dominated by diffusion. In general, 
however, there is a possibility of a rich interplay between the various 
factors which affect the dynamics of the density matrix. Therefore 
it would be worthwhile to do an exact analysis of the master 
equation at absolute zero.\\

\noindent
In Ref. [9], Caldeira and Leggett investigate the destruction of 
coherence between the energy states of an oscillator due to its contact with 
a low temperature environment, in the weak coupling regime. They probe a 
time regime $(\gamma t \ll 1)$ which is complementary to the time regime 
$(\gamma t \gg 1)$ analyzed in the present paper. Their study reveals 
an exponential damping of 
coherence with time. After this initial exponential damping, which lasts 
for a time of the order of $\gamma^{-1}$, there {\it still} remains 
coherence of order unity $(O(1))$ in the system. Then there is an 
intermediate time regime $\gamma t \sim 1$ which is 
not analytically tractable. The present analytical study focuses on the 
late time regime $(\gamma t \gg 1)$ where we find a slow, power law 
decay of coherence. 
This suggests that in the intermediate regime the damping of coherence slows 
down before finally settling down at late times to a power law.\\

\noindent
In Ref. [8], Paz, Habib and Zurek present a detailed analysis of 
a model of a harmonic oscillator linearly coupled to an oscillator bath. 
The results of their study are based on numerical calculations. One of the 
principal aims of Ref. [8] is to test the validity of the {\it high 
temperature} approximation in 
the low temperature regime. They also study the spectral density 
dependence of the decoherence process. This work shows numerically that 
there {\it is} decoherence at absolute zero. The present study goes 
beyond Refs. [8,9] in presenting an analytical expression for the 
temporal decay of coherence at late times.\\

\noindent
{\bf Acknowledgement}\\

\noindent
It is a pleasure to thank N. Kumar, R. Nityananda, J. Samuel, 
Sukanya Sinha and T. Qureshi for illuminating discussions.


\begin{thebibliography}{99}
\bibitem{} J. von Neumann, Mathematische Grundlagen der Quantenmechanik 
(Springer, Berlin, 1932) [English translation by R. T. Beyer, 
Mathematical foundations of quantum mechanics (Princeton Univ. 
Press, Princeton, 1995)]. 
\bibitem{} B. d'Espagnat, Conceptual foundations of quantum mechanics 
(Benjamin, Menlo Park, 1971) pp. 211-214. 
\bibitem{} W.H. Zurek, Complexity, entropy and the physics of information 
(Addison-Wesley, Reading, 1990); Phys. Today 44 (1991) 36, and references 
therein. 
\bibitem{} A. Aspect, P. Grangier and G. Roger, Phys. Rev. Lett. 47 (1982) 
91; \\
A. Aspect, J. Dalibard and G. Roger, Phys. Rev. Lett. 49 (1982) 1804. 
\bibitem{} R.P. Feynman and F.L. Vernon, Ann. Phys. (NY) 24 (1963) 118. 
\bibitem{} A.O. Caldeira and A.J. Leggett, Physica A 121 (1983) 587. 
\bibitem{} W.G. Unruh and W.H. Zurek, Phys. Rev. D 40 (1989) 1071. 
\bibitem{} J.P. Paz, S. Habib and W.H. Zurek, Phys. Rev. D 47 (1993) 488. 
\bibitem{} A.O. Caldeira and A.J. Leggett, Phys. Rev. A 31 (1985) 1059. 
\bibitem{} B.L. Hu, J.P. Paz and Y. Zhang, Phys. Rev. D 45 (1992) 2843. 
\end{thebibliography}
\end{document}